\documentstyle[a4,aps,prb,eqsecnum,graphics]{revtex}

\begin{document}
%\draft
\title{Density-matrix renormalization using three classes of block states}

\author{Marie-Bernadette Lepetit and G. M. Pastor}

\address
{Laboratoire de Physique Quantique, Unit\'e Mixte de Recherche 5626 du CNRS,\\
Universit\'e Paul Sabatier, 118 route de  Narbonne, F-31062 Toulouse, France}

\date{\today}
\maketitle

\begin{abstract}

An extension of the the density matrix renormalization group (DMRG) method 
is presented. Besides the two groups or classes of block states considered 
in White's formulation, the retained $m$ states and the neglected ones, 
we introduce an intermediate group of block states having the following
$p$ largest eigenvalues $\lambda_i$ of the reduced density matrix: 
$\lambda_1 \ge \dots \lambda_m \ge \lambda_{m+1}\ge \dots \ge \lambda_{m+p}$. 
These states are taken into account when they contribute to intrablock 
transitions but are neglected when they participate in more delocalized 
interblock fluctuations.
Applications to one-dimensional models (Heisenberg, Hubbard and
dimerized tight-binding) show that in this way the involved computer 
resources can be reduced without significant loss of accuracy. 
The efficiency and accuracy of the method is analyzed by varying $m$ 
and $p$ and by comparison with standard DMRG calculations. 
A Hamiltonian-independent scheme for choosing $m$ and $p$ and for 
extrapolating to the limit where $m$ and $p$ are infinite is provided.
Finally, an extension of the 3-classes approach is outlined, which 
incorporates the fluctuations between the $p$ states 
of different blocks as a self-consistent dressing of the block 
interactions among the retained $m$ states. 

\end{abstract}
%
%\vspace{0.5cm}
%
%71.10.Fd  Lattice fermion models (Hubbard model, etc.)
%02.70.-c  Computational techniques
%75.10.Jm  Quantized spin models 

%\narrowtext

\section{Introduction}
\label{sec:Intro}

The density matrix renormalization group (DMRG) method introduced 
by S.~R.~White in 1992 \cite{dmrg0} has rapidly become
one of the leading numerical tools for the study of one-dimensional (1D) 
and quasi-1D strongly correlated systems. Accuracy and flexibility are
two main qualities of this approach. Indeed, the DMRG method allows to 
study large many-body systems containing hundreds of sites and to 
determine several ground-state and low-lying excited-state properties 
with a precision nearly comparable to that of conventional exact 
diagonalization methods. The success of this technique is readily demonstrated
by its wide range of applications. These concern Heisenberg spin
Hamiltonians \cite{dmrg0,dmrg-heis}, the $t$-$J$ model \cite{dmrg-tJ}, 
Hubbard-like Hamiltonians on various lattices (e.g., dimerized polymer
chains and Bethe lattices) \cite{dmrg-dim,pang,toi-polyac,cousy}, 
the Kondo impurity problem \cite{dmrg-kondo}, 
acoustic phonons \cite{dmrg-phonons}, etc.  

In its simplest form, the method can be regarded as an iterative
projection technique, which allows to include the most relevant part
of the ground-state wave-function on a limited number of many-body
states and which leads to renormalized block interactions between
different regions in real space.  The accuracy of the calculation 
is mainly controlled by the number $m$ of basis states which are 
retained in each
block after a renormalization iteration. When combined with efficient
diagonalization techniques for sparse matrices (e.g., Lanczos or
Davidson methods) the ground-state energy of an infinite 1D system can
be extrapolated with a remarkable precision at a much lower
computational cost than conventional exact diagonalizations (e.g., 
straightforward Lanczos).
However, as the systems under study become more complex (e.g., itinerant 
electrons in rings, ladders or multi-band wires) and as more delicate 
properties are considered (e.g, charge and spin gaps, dynamical response 
functions \cite{dmrg-dyn}, etc.) the DMRG method faces its own 
limitations, particularly 
since accuracy improves slowly with increasing $m$. In such cases
the value of $m$ and thus the size of the projected Hilbert space 
required for an appropriate accuracy become excessively large. The
involved computer resources exceed reasonable limits and in many relevant 
situations the calculations are simply non-feasible. 
It is therefore of considerable interest to improve the efficiency 
of the DMRG method in order to reduce the computational effort without 
significant loss of accuracy.

In this paper an extension of DMRG method is presented which 
takes better profit from the hierarchy among the states spanning 
the renormalized blocks. As proposed by White in his original 
paper \cite{dmrg0} each block $b$ is best described in terms of the 
eigenstates $\vert i\rangle_b$ of the reduced 
ground-state density matrix. The corresponding eigenvalues $\lambda_i$
represent the population of the block-state $\vert i\rangle_b$ 
and thus provide a natural hierarchy among them. Two classes of states 
are considered in the original DMRG procedure: the $m$ states having 
the largest $\lambda_i$ which are retained and the remaining states
which are completely neglected. In this paper a larger flexibility
is introduced by distinguishing an additional group of $p$ states
($\lambda_1 \ge \dots \ge \lambda_m \ge \lambda_{m+1} \ge\dots\lambda_{m+p}$)
which are primarily important for the description of the ground-state 
wave function at block $b$ and which can be neglected when they participate 
to long-range interblock fluctuations.
It is shown that in this way the computer resources involved in 
the calculations can be appreciably reduced without introducing 
significant additional inaccuracies. 
Moreover, it is also interesting to vary the values of $m$ and $p$ 
and to quantify how this reflects in the physical properties in order to
gain further insight into the internal many-body structure 
of the renormalized superblocks.

The rest of the paper is organized as follows. In the next section 
the 3-classes renormalization method is presented and compared
with White's original approach \cite{dmrg0}. In Sec.~\ref{sec:results} 
the method is applied to a few relevant models for which exact or 
good numerical results are available. Improvements and limitations
of the 3-classes method are discussed in particular by analyzing 
the results as a function of the sizes $m$ and $p$ of class-1 and 
class-2 subspaces. A model independent scheme for optimizing the choice of
$m$ and $p$ and for extrapolating to the limit where $m$ and $p$ are 
infinite is also considered. Finally, Sec.~\ref{sec:conc} summarizes 
our conclusions and outlines a self-consistent extension of the present 
approach.

\section{3 classes DMRG method}
\label{sec:method}

In this section we present a density-matrix renormalization method 
using 3 classes of block states. First of all, White's original DMRG 
algorithm for infinite one-dimensional systems \cite{dmrg0} 
is briefly recalled for the sake of clarity and in order to facilitate 
comparisons. 

\begin{figure}[x]
\centerline{\resizebox{9cm}{2.5cm}{\includegraphics{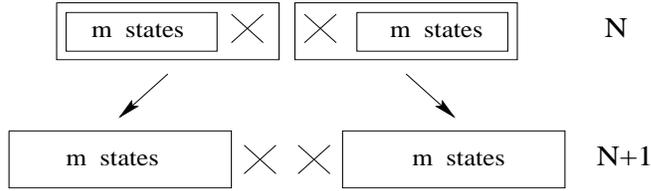}}}
\vspace{0.5cm}
\caption{
Illustration of the superblock renormalization for infinite one-dimensional
systems. Crosses represents atomic sites and rectangles
renormalized blocks. Notice that the total number of sites is increased 
by 2 at each renormalization iteration $N$. 
        }
\vspace{1cm}
\label{fig:ren}
\end{figure}

The DMRG method is an iterative procedure. In the case of infinite open 
chains the ground-state properties of the infinite system are obtained 
from the extrapolation of a succession of calculations on finite-length 
chains. At each renormalization
group (RG) iteration $N$, the number of sites $L$ is increased by 2
($L = 2N+2$).  In spite of the fact that $L$ increases with $N$, 
the dimension of the Hilbert space is kept constant by means of the
following approximation. At each RG iteration the Hilbert space of the
$L$-site chain is projected onto a subspace ${\cal E}$
spanned by a limited number of many-body states $\Psi_{ijkl}$
which are intended to take into account the most relevant part of the 
ground-state wave function. The $\Psi_{ijkl}$ are constructed as the
antisymmetrized direct product of many-body states, which are usually 
referred to as ``block states.'' These blocks correspond to different 
regions in real space as illustrated 
in Fig.~\ref{fig:ren}. 
Let $\psi^L_i$ ($\psi^R_i$) stand for $i$th
many-body state of the left (right) block ($1\le i \le m$) and let $\phi_j$
be the $j$th state of a central (unrenormalized) site ($1\le j\le m_1$). 
Then
\begin{equation}
\Psi_{ijkl} = \psi^L_i \otimes   \phi_j \otimes \phi_k \otimes \psi^R_l \; . 
\label{eq:Psi}
\end{equation}
White showed that the error introduced by the truncation of the
Hilbert space is minimized when the states spanning the subspace of 
each renormalized block $b$ are the eigenvectors corresponding to 
the $m$ largest eigenvalues of the ground-state density matrix reduced 
to $b$. In other words, only the block states having
the largest occupations are kept.  

The DMRG algorithm can be
summarized by four main steps involved in going from iteration $N$ to 
iteration $N+1$: 
{\em i)} The ground state of the $L$-site system ($L=2N+2$), 
which corresponds to the Hilbert space ${\cal E}(N)$ of the 
iteration $N$, is determined. This involves 
a calculation in a space of limited dimension $(m m_1)^2$. 
Sparse-matrix diagonalization methods (e.g., Lanczos or
Davidson algorithm) are normally used. 
{\em ii)} The reduced ground-state density matrices $\rho^L$ and $\rho^R$
of the superblocks are calculated. As illustrated in Fig.~\ref{fig:ren},
the superblocks are formed by one of the side blocks ($L$ or $R$) and its
neighboring site. The eigenvalues $\lambda^L_i$ ($\lambda^R_i$)
and eigenvectors $\psi^L_i$ ($\psi^R_i$) of $\rho^L$ ($\rho^R$) 
are determined. 
{\em iii)} The superblock subspace of dimension
$m m_1$ is projected onto the $m$ most populated states derived 
in the previous step ($\lambda_1\ge \dots \ge\lambda_m$). These are 
the left and right block-states of the next iteration. In this way the 
central sites have been projected or renormalized into the side blocks.
{\em iv)} The Hilbert space ${\cal E}(N+1)$ of the iteration $N+1$,
which corresponds to a system with $2N+4$ sites, is constructed according 
to Eq.~(\ref{eq:Psi}). The renormalized Hamiltonian is determined 
by performing the appropriate unitary 
transformation and projection. Finally, the procedure loops back to step
{\em i)} until the physical properties extrapolated to $L=\infty$ converge.

Two groups or {\em classes} of block states are distinguished in 
White's DMRG procedure. The first one is given by the $m$ most 
populated states which are retained, and the second one by the remaining
$m m_1 - m$ least populated states which are projected out. 
This criterion certainly respects the hierarchy of states given by the 
eigenvalues or 
populations of the different block states. However, it results in a rather 
rigid procedure, since the contributions of a block state must be 
either kept in full or completely neglected. Except in cases where 
the number of states that can be kept is large enough to ensure the 
desired accuracy or when a systematic extrapolation to $m=\infty$ 
can be performed, the choice of $m$ is a delicate and drastic one. 
In general there is no clear gap in the populations of the block
states which could justify where the truncation should take 
place. It is our purpose to render the method more flexible 
and efficient by the introduction of a new intermediate class of block 
states which are intended to contribute in a restricted way 
to the construction of the Hilbert space of the total system.
Physically, one expects that the description of the internal degrees 
of freedom of the superblocks and the renormalization of block-site 
interactions should be more demanding and more relevant to the 
ground-state properties than the description of fluctuations 
between left and right block-states which are farther apart.
The intermediate class of states should take this into account.
Moreover, with three classes of block states, the
transition between kept and discarded many-body processes
should be smoother and the actual choice of $m$ less critical.

\begin{figure}[x]
\centerline{\resizebox{12cm}{7cm}{\includegraphics{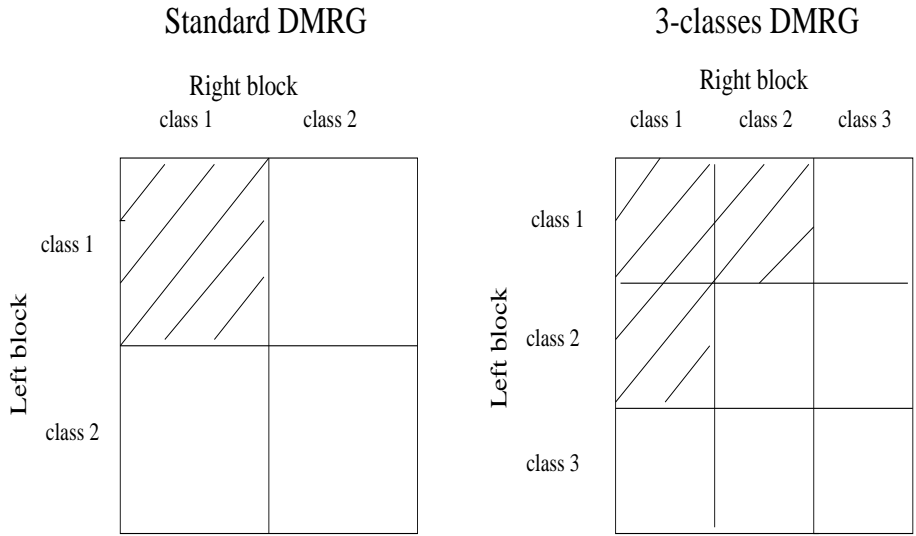}}}
\vspace{0.5cm}
\caption{
Illustration of the different projection subspaces  ${\cal S}$
in White's DMRG method 
\protect\cite{dmrg0} and in the present 3-classes renormalization. 
The complete Hilbert space ${\cal E}$ involves all the antisymmetrized 
direct products of left and right block-states 
(see Fig.~\protect\ref{fig:ren}).
The hashed areas indicate the subspaces ${\cal S}$ which are kept 
at each iteration.
        }
\vspace{1cm}
\label{fig:st}
\end{figure}

We consider 3 groups or classes of eigenstates of the block
density matrix: {\em i)} The first group (class 1) is given by
the $m$ most populated states $\psi_i$ where $1\le i \le m$ and
$\lambda_1\ge \dots\ge\lambda_m$. {\em ii)} The second group 
(class 2) contains the eigenstates $\psi_i$ associated to the following
$p$ eigenvalues, i.e., $m+1 \le i \le m+p$ with 
$\lambda_m\ge \lambda_{m+1} \ge \dots\ge\lambda_{m+p}$.
{\em iii)} Finally, the third group (class 3) refers to the remaining
$m m_1 - (m +p)$ states.
The states of the first and third classes are treated just as in 
White's algorithm, i.e., all direct-product states or configurations of
the complete system $\Psi_{ijkl}$ which involve only class-1 block-states 
($i,l \le m$) are taken into account while configurations involving a
class-3 state ($i$ or $l > m+p$) are all projected out 
[see Eq.~(\ref{eq:Psi})].
An intermediate criterion is used for the second class,
namely, class-2 states of the left superblock are included only
in combination with class-1 states of the right superblock and vice versa.
In other words, direct-product wave-functions $\Psi_{ijkl}$ 
with $m+1\le i\le m+p$ and  $m+1\le l\le m+p$ are neglected.
This is illustrated in Fig.~\ref{fig:st}.

In analogy with configurations interaction methods, the subspace
spanned by the set of $\Psi_{ijkl}$ with $i,l \le m$ can be regarded
as a complete active space (CAS), the subspace spanned by the 
$\Psi_{ijkl}$ with 
$i \le m$ and $m+1 \le l \le m+p$ or 
$l \le m$ and $m+1 \le i \le m+p$ defines then a kind of mono-excited space,
and the discarded states obtained from $\Psi_{ijkl}$ 
with $m+1 \le i,l \le m+p$ can be viewed as double many-body excitations. 
Class-2 states have a relatively small weight in the
ground-state wave-function. One therefore expects that with an 
appropriate choice of $m$ and $p$ the contributions of $\Psi_{ijkl}$ 
obtained by the tensor product of two such block states should 
be particularly small, eventually even less important than the states 
involving products of a class-1 and class-3 states which are 
already discarded in White's DMRG method.
Therefore, the accuracy of a 3-classes renormalization should not differ 
significantly from that of a standard 2-class DMRG-calculations retaining 
$m+p$ states. 

Depending on the values of $m$ and $p$, the 3-classes 
procedure may allow an appreciable reduction of the dimension 
of the many-body Hilbert space, namely, from $D_0 = m_1^2(m+p)^2$ 
to $D = m_1^2(m^2 + 2mp)$.
This is of considerable interest in many practical applications which 
are often limited by computer-memory needs and most critically by the computer 
time required for the determination of the ground-state wave-function 
at each RG iteration.
The 3-classes approach can also be applied straightforwardly to more
complex lattices such as Bethe lattices, which involve renormalizations 
at more than two blocks \cite{cousy}. In these cases the reduction of 
computational effort is even more important, since 
$D_0/D = (m+p)^{n_b}/ (m^{n_b} + n_b\;\! p\;\! m^{n_b-1})$ where $n_b$ is 
the number of renormalized blocks (e.g., $n_b = z$ for a Bethe lattice 
with $z$ nearest neighbors). 

The eigenvalues $\lambda_i$ of the density matrix reduced to block $b$ 
represent the probability of block $b$ being in state $\psi_i^b$ 
($b= L$ or $R$). Therefore, the traces  
$\Sigma_1= \sum_{i=1}^m \lambda_i$, $\Sigma_2= \sum_{i=m+1}^{m+p} \lambda_i$
and $\Sigma_3= \sum_{i=m+p+1}^{m_1 m} \lambda_i$, measure the weight of 
each of the 3 classes of block states on the ground-state wave function
($\Sigma_1 + \Sigma_2 + \Sigma_3 = 1$). 
Approximating for simplicity the norm of the retained part of the system 
wave-function by the product of the traces $\Sigma_i$ on the different 
blocks and keeping only the leading terms in $\Sigma_2$ and
$\Sigma_3$ ($\Sigma_2, \Sigma_3 \ll 1$), one estimates the truncation 
error $\epsilon(m,p)$ involved in a 3-classes renormalization as
$\epsilon(m,p) = n_b \Sigma_3 + ({n_b \atop 2}) \Sigma_2^2$, where  
$n_b$ refers to the number of renormalized blocks ($n_b = 2$ in the present 
calculations). If $p=0$, $\Sigma_2 = 0$ and we recover White's estimation 
of the truncation error $1-P_m$ which is proportional to 
$\Sigma_3$ \cite{dmrg0} ($\Sigma_3 = 1 -\Sigma_1 = 1 -P_m$ if $\Sigma_2 = 0$).
As discussed in Sec.~\ref{sec:results}, comparison with exact results 
show that $\epsilon(m,p)$ gives very good estimation of the errors
involved in 3-classes calculations. In practice,  $\epsilon(m,p)$ can be 
used either to control the accuracy of the results, which often depend 
on the model Hamiltonian and its parameters, or to extrapolate to the 
limit $\epsilon(m,p)\to 0$ ($m,p \to \infty$) in order to further reduce 
the errors. A similar procedure has been already applied in the 
standard DMRG \cite{dmrg0}.
Moreover, $\epsilon(m,p)$ provides a practical criterion for optimizing 
the value of $m$ and $p$, for instance, when the size of the Hilbert 
space $D(m,p)$ is fixed. In general it is a good choice to keep 
$\Sigma_2^2$ and $2\Sigma_3$ of the same order of magnitude. In fact, 
for a fixed $D(m,p)$ a reduction of $\Sigma_2^2$ (e.g, by decreasing $p$) 
implies an increase of $\Sigma_3$ and vice versa.

\section{Results}
\label{sec:results}

The density-matrix renormalization using 3-classes of states has 
been applied to several benchmark problems in order to check the accuracy 
of the method and to analyze the quality of the results as a 
function of the subspace dimensions $m$ and $p$. In particular
comparison is made with the standard DMRG algorithm \cite{dmrg0}. 
The considered systems include the spin-$1/2$ Heisenberg chain, the periodic
Hubbard chain as a function of correlation strength $U/t$,
and the dimerization of polyacetylene using a distance-dependent 
tight-binding Hamiltonian. In the following our main results are discussed.

\subsection{Spin-$1/2$ Heisenberg chain}          
\label{sec:heis}

In this section we consider the infinite spin-$1/2$ Heisenberg chain 
as a representative application of the 3-classes DMRG method 
to spin systems. This problem is particularly interesting 
from a methodological point of view since the exact Bethe-Ansatz
solution as well as extremely good standard DMRG calculations are 
available \cite{dmrg0}. In Table \ref{tab:Heis} our results 
for the ground-state energy $E$ are compared with the exact 
value $E_{ex} = -J \ln 2 + J/4$ and with DMRG calculations using White's
method which corresponds to $p=0$. In all cases, $N=100$ renormalization 
iterations are performed. This ensures that the extrapolated ground-state 
energy of the infinite-length chain is converged at least within $10^{-6} J$.
Small values of $m$ and $p$ are considered in Table \ref{tab:Heis}
in order to compare the convergence and performance of the different 
methods. Notice, however, that more accurate calculations involving 
a larger number of states (e.g., $m=m+p=44$ \cite{dmrg0}) are not 
at all demanding with present computer facilities. For $m$ and $p$ larger
than the values shown in Table \ref{tab:Heis}, the 2-classes and 
3-classes DMRG results converge very well to the exact
ground-state energy and are almost indistinguishable from each other. 

The 3-classes results recover accurately the exact ground-state energy 
even if as few as $5$ states are kept in the class-1 subspace or complete 
active space (CAS). For $m=5$ and $m+p = 21$, $E-E_{ex}\simeq 10^{-4}J$,
which seems an acceptable accuracy taken into account that the size of 
the Hilbert space is reduced significantly ($D_0/D = 2.4$). Our results 
show that the intrablock fluctuations between the $m$ class-1 states and 
the $p$ class-2 states play a major role in the 
description of the ground-state wave function. In fact, neglecting them 
yields so poor results that for $m=m+p=5$ the DMRG procedure 
diverges after a few iterations. 
In contrast, the effects of quantum fluctuations between class-2 states
of different blocks are much less important for the determination 
for the ground-state energy. These contributions
are neglected in a 3-classes DMRG calculation with $m=5$ ($m=7$) and $m+p=21$ 
which yields $E-E_{ex}= 1.07\times 10^{-4}J$ 
($E-E_{ex}= 4.47 \times 10^{-5}J$) 
and are included in the calculation with $m=m+p=21$ which yields 
$E-E_{ex}= 2.55 \times 10^{-5}J$. Similar conclusions are derived 
for other values of $m$ and $p$.

The calculations for different $m$ and fixed $m+p$ provide further insight 
on the convergence properties of the method. For very small $m$ the 
difference between 3-classes and 2-classes results decreases
very rapidly for increasing $m$ (see Table~\ref{tab:Heis} for $m=5$ and $7$). 
This indicates that a minimum size of the class-1 subspace is 
indispensable for obtaining good accuracy. Further increase of $m$
yields almost the same accuracy as the calculation with $m=m+p=21$. 
Nevertheless, the reduction of the dimension of the Hilbert space is 
still appreciable (e.g., $D_0/D = 1.4$ for $m=10$).
Let us recall that the error in $E$ decreases slowly with 
increasing $D$ already in the standard DMRG method \cite{dmrg0}. 
As in the case of very small $m$, the intrablock transitions involving 
class-2 states are very important for improving the accuracy 
($E-E_{ex} = 2.06\times 10^{-4} J$ for $m=m+p=10$). 
We conclude that the 3-classes DMRG procedure improves efficiency without 
significant loss of accuracy, provided that the size $m$ of the CAS is 
not too small ($m\ge 7$ in the present case). 

The truncation error $\epsilon(m,p) = \Sigma_2^2 + 2\Sigma_3$ 
provides a very good estimate of the accuracy of the calculations 
for all considered values of $m$ and $p$. In fact, $E-E_{ex}$ follows 
very closely a linear relation with $\epsilon(m,p)$. It is therefore 
simple to extrapolate to the limit $\epsilon(m,p) \to 0$ in order to 
further improve the results (see Table~\ref{tab:Heis}). A similar procedure 
is used in standard DMRG \cite{dmrg0}. Moreover, the 
results suggest that the dimensions of class-1 and class-2 
subspaces could be chosen in order to optimize accuracy by minimizing 
 $\epsilon(m,p)$ for a given computational cost.

\subsection{Hubbard chain}

The DMRG method converges very rapidly and accurately when applied to
problems involving strong electron correlations such as the Heisenberg 
model. However, the performance of the method is very sensitive to the 
strength of the Coulomb interactions and to the degree of electron 
delocalization. For instance, in the case of the Hubbard chain at 
half-band filling and for a given number of block states $m$, the relative 
error $\Delta\varepsilon$ in the ground-state energy increases by an order 
of magnitude as we move from the strongly correlated to the uncorrelated 
limit. For example for $m=80$ ($p=0$), 
$\Delta\varepsilon = 8.1 \times 10^{-5}$ for $U/t = 76$, 
$\Delta\varepsilon = 4.9 \times 10^{-5}$ for $U/t = 4$,
$\Delta\varepsilon = 3.1 \times 10^{-4}$ for $U/t = 1$ and 
$\Delta\varepsilon = 3.3 \times 10^{-4}$ for $U/t = 0$. 
Moderate and weak electron interactions are thus much more difficult 
to describe with the DMRG method. In addition, for small $U/t$ the 
algorithm requires a quite larger number $N$ of RG steps to converge. 
Typically, $N$ is about $2$--$3$ times larger for $U=0$ than for 
very large $U/t$ \cite{toi-polyac}. It is therefore considerably interesting 
to analyze the accuracy and convergence properties of the 3-classes 
DMRG method in applications to itinerant-electron models such as 
the Hubbard model.

\begin{figure}[x] 
\centerline{\resizebox{8cm}{16cm}{\includegraphics{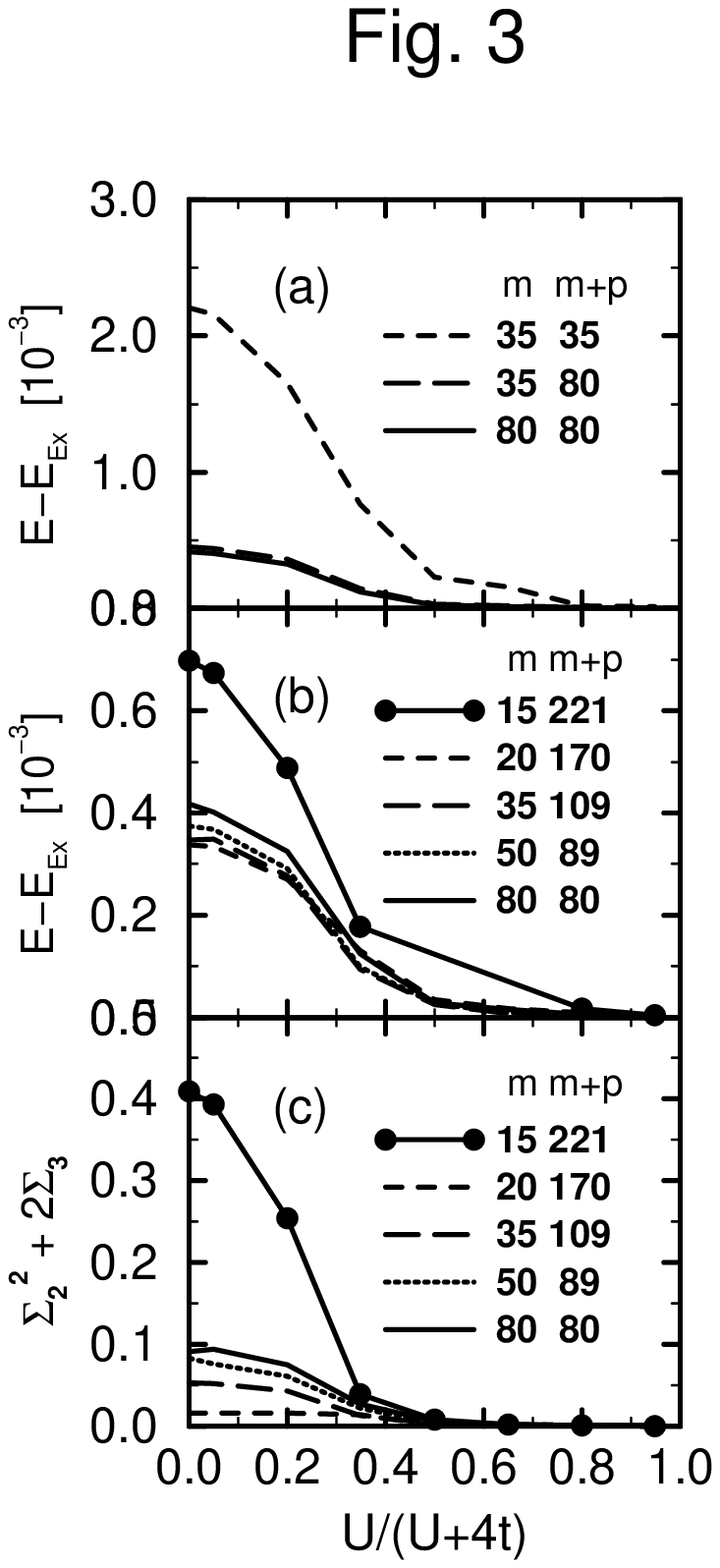}}}
\vspace{0.5cm}
\caption{
DMRG results for the ground-state energy $E$ of the one-dimensional 
Hubbard model at half-band filling as a function of the Coulomb repulsion 
strength $U/t$. $E_{ex}$ refers to the Bethe-ansatz exact  
solution \protect\cite{lieb}. The number $m$ ($p$) of block states in the 
class-1 (class-2) subspaces is indicated (see Fig.~\protect\ref{fig:st}). 
White's DMRG method corresponds to $p=0$. 
In (a) the dimensions $D(m,m+p)= 16[(m+p)^2 - p^2]$ of the many-body 
Hilbert spaces are $D(35,35)= 19600$, $D(35,80)= 70000$, $D(80,80)= 102400$.
In (b) $m$ and $p$ are varied keeping a constant $D(m,m+p)\simeq D(80,80)$. 
In (c) the estimation of the truncation error 
$\epsilon(m,p) = \Sigma_2^2 + 2 \Sigma_3$ is shown.
        }
\vspace{1cm}
\label{fig:hub}
\end{figure}

In Fig.~\ref{fig:hub} results are given for the ground-state energy of 
the half-filled one-dimensional Hubbard chain as function of the Coulomb 
repulsion strength $U/t$. These were obtained by using the 3-classes 
DMRG method with various representative values of $m$ and $p$. 
The standard DMRG algorithm corresponds to $p=0$. In all cases, the number 
of renormalization iterations $N$ is increased until the infinite-length 
extrapolated energy has converged within an error smaller
than $10^{-6}t$. Starting from the uncorrelated limit, the accuracy 
improves considerably with increasing $U/t$. Very precise results 
are obtained particularly for $U/t>4$. However, for very large $U/t$ 
($U/t > 16$) the relative error may increase slightly. 
For example, for $U/t = 76$ ($U/t = 16$) the relative errors 
$\Delta\varepsilon = (E-E_{ex})/E_{ex}$ are 
$\Delta\varepsilon =4.2 \times 10^{-4}$ 
      ($\Delta\varepsilon =1.1 \times 10^{-4}$) for $m=m+p=35$,
$\Delta\varepsilon =8.0 \times 10^{-5}$ 
      ($\Delta\varepsilon =3.5 \times 10^{-5}$) for $m=35$ and $m+p=80$, and
$\Delta\varepsilon =8.0 \times 10^{-5}$ 
      ($\Delta\varepsilon =3.0 \times 10^{-5}$) for $m=m+p=80$.
Notice that, although the differences between the calculation with 
$m=m+p=35$ and with $m=m+p=80$ are significant, the 3-classes results 
follow very closely the $m=m+p=80$ curve.
This implies that the largest part of the improvements made by 
increasing $m$ from $35$ to $80$ are recovered by taking $m=35$ and 
$m+p=80$, i.e., by including the intrablock fluctuations. Transitions
involving class-2 states on two different superblocks 
are not very important in this case. This allows to reduce the dimension of 
the Hilbert space by about 30\%. The results seem remarkable, particularly 
if we recall that the 3-classes method is a controlled approximation in 
the sense that it respects the variational principle. At each RG 
iteration the energy obtained with $m=35$ and $m+p=80$ is an upper 
bound for the energy of a standard DMRG calculation with $m=80$ and $p=0$.

The same trends hold for weaker interactions. For $U<4t$, one observes 
that the differences between the relative errors in the 3-classes and 
in the standard DMRG calculations increase. At the same time, however, 
the $m=m+p=80$ curve starts to deviate more significantly from the exact 
result showing the that block states $i$ with smaller eigenvalues of the 
density matrix (i.e., weaker occupations $\lambda_i < \lambda_{80}$) should 
be taken into account. Quantitatively, neglecting the intrablock fluctuations
involving states with $\lambda_i < \lambda_{80}$ is a much more important 
source of error than the interblock fluctuations neglected in the 3-classes
method [see Fig.~\ref{fig:hub}(a)]. The additional error introduced by the
use of the 3-classes procedure ($m=35$ and $m+p =80$) is only 9\% of the 
total error ($m=m+p =80$). Therefore, in order to 
further improve accuracy for $U/t\le 4$ it should be more efficient to 
increase the size $p$ of the class-2 subspace instead of increasing $m$ 
with $p=0$ as in the standard DMRG method. This is consistent with our
estimation of truncation error which is quadratic in the sum of the 
class-2 eigenvalues $\Sigma_2$ and linear in $\Sigma_3$. 

In order to investigate this question we compare in Fig.~\ref{fig:hub}(b) 
the accuracy of the 3-classes approach as a function of the size of
class-1 and class-2 subspaces. Several values of $m$ and $p$ are 
considered which all correspond to approximately the same dimension 
of the Hilbert space of the complete system:
$D(m,m+p)= 16[(m+p)^2 - p^2] \simeq  D(80,80) = 102400$.
Therefore, all these calculations demand essentially the same computer
resources. One observes that there is an optimum choice 
for $m$ and $p$ for which the relative error $(E-E_{ex})/E_{ex}$ 
is reduced by about 20--30\% with respect to the standard calculation 
with $m=m+p=80$. As in the Heisenberg model, a minimum number of class-1 
states is indispensable for describing 
the ground-state of the entire system with a reasonable accuracy.
If $m$ is too small, for instance $m=15$, the results are poorer
even if $p$ is very large. Once the most important configurations are 
taken into account as class-1 states, it is more useful to maximize $p$ 
in order to increase the flexibility of the wave-function within each 
superblock rather than to increase the number of class-1 states. This 
improves the treatment of intrablock correlations and allows a more 
precise renormalization of the interactions as the number of sites 
grows. The results shown in Fig.~\ref{fig:hub}(b) for $m=15$ ($p=221$), 
$m=20$ ($p=170$) and  $m=35$ ($p=109$) illustrate this behavior.

Taking into account the strong dependence of $E-E_{ex}$ on $(m,p)$ 
and on $U/t$ (even for $p=0$) it is of considerable practical interest
to have a model-independent criterion for choosing $m$ and $p$
and for controlling the accuracy of 3-classes DMRG calculations.
Comparison between Figs.~3(b) and 3(c) shows that the estimation of 
the truncation error $\epsilon(m,p)=\Sigma_2^2 + 2\Sigma_3$ follows 
the trends in $E-E_{ex}$. Therefore, the minimization of 
$\epsilon(m,p)$ provides an appropriate means of choosing $m$ and $p$
for a given dimension of the Hilbert space. Alternatively, one may
use $m$ and $p$ such that $\epsilon(m,p)$ is constant in
order to ensure approximately the same accuracy in
different calculations (e.g., as a function of $U/t$ or of the 
dimerization $\delta$). Notice, that $\epsilon(m,p)$ can be easily 
computed at each RG iteration.

It is should be also noted that the minimum $m$, which yields
better accuracy than the standard DMRG method, as well as the optimal 
choice for $m$ and $p$ for a given $D$ depend on the Coulomb interaction 
strength $U/t$. 
For example, $m= 20$ and $p=170$ yields the best results for $U/t\le 1$, 
but for $U/t\ge 2$ this is somewhat less accurate than the standard DMRG 
calculation. In the range $0\le U/t \le 4$ the optimum $m$ ($p$) 
increases (decreases) as $U/t$ increases. This is the result of two
contributions to the minimization of $\Delta\varepsilon$, which compete 
when the dimension of the Hilbert space is fixed. Increasing $p$  
improves the description of the wave function within each block as well 
as the superblock renormalizations since it reduces $\Sigma_3$. However, 
this is done at the expense of decreasing $m$ (or increasing $\Sigma_2$)
and therefore tends to worsen the description of the
interactions between different blocks.  
For $U/t>4$ the results for $m=35$--$80$ ($p=109$--$80$) are very similar. 
Close to the optimum $(m,p)$ the energy differences remain small.
Let us finally recall that the overall convergence and accuracy of 
DMRG methods improve as local charge fluctuations are reduced. 
The quantitative differences between different renormalization or 
projection strategies become therefore less important at large $U/t$. 

\subsection{Dimerized chains}
\label{sec:poly}

The purpose of this section is to discuss the application of the 3-classes
DMRG method to dimerized polymer chains. The dimerization of 
polyacetylene is particularly interesting since it depends sensitively
on the details of the wave function and thus on the number of states
$m$ used in DMRG calculations \cite{toi-polyac}. A distance dependent
tight-binding Hamiltonian is considered for the $\pi$ valence 
electrons in polyacetylene. Repulsive interactions between $\sigma$ 
electrons are modeled by a pairwise potential 
$E_\sigma(r_{ij})$ \cite{toi-polyac}. 
The hopping integrals $t(r_{ij})$ and the repulsive potential 
$E_\sigma(r_{ij})$ are obtained from {\em ab-initio} calculations
on the ground-state energy and first singlet-triplet excitation 
energy of the ethylene molecule \cite{daniel}.
The main interest of the uncorrelated case is methodological, 
since this is the most difficult limit to study using the DMRG method. 
DMRG calculations for realistic finite values of the Coulomb repulsion $U$ 
are reported in Ref.~\cite{toi-polyac}.

\begin{figure}[x] 
\centerline{\resizebox{12cm}{12cm}{\includegraphics{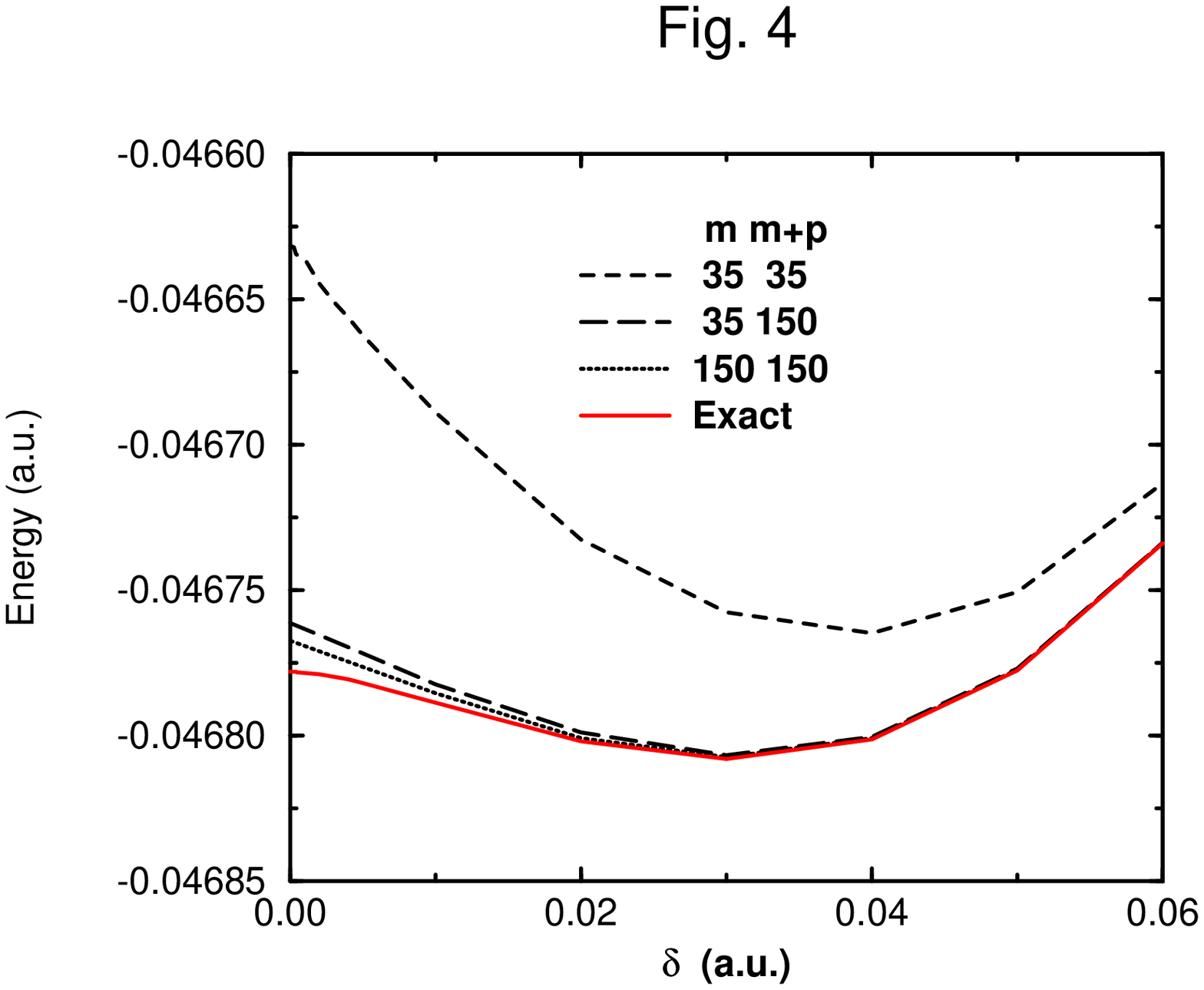}}}
\vspace{0.5cm}
\caption{Ground-state energy $E$ of polyacetylene infinite chains 
(in a.u.) as a function of the dimerization 
$\delta=\vert r_{i,i+1} - r_{i-1,i}\vert /2$ 
for $R = (r_{i,i+1} + r_{i-1,i})/2 = 2.66 a_0$. 
The results are obtained using a distance-dependent tight-binding model 
($U=0$) which parameters are derived from {\em ab initio} calculations
on the ethylene molecule \protect\cite{daniel}. 
The number of block states $m$ and $p$ kept in the 3-classes
DMRG calculations are indicated in the inset. 
The exact tight-binding result is given by the lowest thin curve.
        }
\label{fig:poly}
\end{figure}

In Fig.~\ref{fig:poly}
results are given for the ground-state energy $E(\delta)$ of an 
infinite polyacetylene chain vs.\ the dimerization 
$\delta = \vert r_{i,i+1} - r_{i-1,i} \vert / 2$, as obtained using the 
model Hamiltonian described above. The exact tight-binding
result is also given for the sake of comparison. As in the undimerized 
Hubbard chain, the 3-classes calculations with $m=35$ and $m+p = 150$
remove the largest part of the discrepancies between the $m=m+p=35$ and
$m=m+p=150$ standard calculations. In particular the position of the 
minimum $\delta_{min}$ is improved considerably. However, notice that 
the error $\Delta E = E-E_{ex}$ decreases with increasing $\delta$
and therefore, $\delta_{min}$ remains overestimated. For $m=m+p=35$ 
($m=m+p=150$) one obtains $\delta_{min}=0.389 a_0$ 
($\delta_{min}=0.306 a_0$). Using $m=35$ and $m+p = 150$ one finds
$\delta_{min}=0.310 a_0$, while the exact result is 
$\delta_{min}=0.303 a_0$. As expected, for a given $m$ and $p$, 
the 3-classes method converges better towards the more demanding 2-classes 
calculation ($m+p$ states per block) as the dimerization increases.

\section{Summary and Outlook}
\label{sec:conc}

In this paper, an extension of the the density matrix renormalization
group (DMRG) method has been presented, which achieves larger
flexibility and improved performances without increasing the
computational effort. An intermediate class of block states is
introduced between the states which are fully taken account and those
which are discarded at each renormalization step. These states
contribute only as far as {\em intra}block fluctuations are concerned
and provide a smoother criterion for discerning between retained and
discarded many-body states. Applications of the method to
one-dimensional models (Heisenberg, Hubbard and dimerized
tight-binding) show that in this way the involved computer resources
can be reduced without significant loss of accuracy. Varying the
number $p$ of intermediate states relative to the number $m$ of block
states which are fully retained one obtains further insight into the
DMRG method, particularly concerning the
internal structure of the renormalized blocks and the relative importance
of intrablock and interblock fluctuations. For example, for a
given dimension $D(m,p)$ of the Hilbert space of the renormalized
system, optimum values of $m$ and $p$ can be determined which minimize
the error in the ground-state energy. In these cases the discrepancies
with exact results are 20--30\% smaller than in standard DMRG
calculations. Using a simple estimation of the truncation error
involved in a 3-classes renormalization, a model-independent strategy 
is derived for choosing $m$ and $p$ and for extrapolating to the limit 
where $m$ and $p$ are infinite. In this way reliability and accuracy are
further improved. 

In order to pursue the development of the 3-classes approach, the 
fluctuations involving class-2 states of different blocks should be
taken into account without increasing the dimension of the renormalized 
Hilbert space. This can be done by using a self-consistent intermediate
Hamiltonian method based on a coupled-cluster approximation. Similar
techniques are very successful in large-scale {\em ab initio} configurations
interaction (CI) calculations on molecular systems \cite{intham}. The
model space ${\cal S}$ of the effective Hamiltonian coincides with the
Hilbert space of the 3-classes DMRG approach (i.e., multiple class-2
states on different blocks are excluded). From the calculation of the
ground-state wave function $\vert\Psi\rangle$ within ${\cal S}$ one
obtains relations between the wave-function coefficients corresponding
to class-1 and class-2 states.  Assuming a coupled-cluster expansion
for $\vert\Psi\rangle$, these relations allow to infer approximately
the coefficients of the wave function beyond ${\cal S}$, i.e., when
class-2 states are present on different blocks.  These fluctuation
effects can be then included as a dressing of the block interactions
within ${\cal S}$. Since the dressed or intermediate Hamiltonian
depends on $\vert\Psi\rangle$, a set of self-consistent equations must
be solved at each renormalization iteration. Notice, however, that
this could be done within the Lanczos or Davidson iterative procedure
in order to avoid a too large increase of computer time. The
detailed formulation of this self-consistent approach is rather
lengthy and beyond the scope of this paper. A summary may be found in
the Appendix. While an efficient implementation of such an extension
should further improve performance, it is also true that its complexity 
risks to limit possible applications to very 
difficult cases where a sufficiently large number of states $m$ and $p$ 
is not practicable or where a systematic extrapolation to the limit where 
$m$ and $p$ are infinite is not possible.

\acknowledgements 
Computer resources provided by IDRIS under project 960806 are
gratefully acknowledged.

\section*{Appendix} 

The aim of this section is to outline a method for taking into account
the effect of the neglected fluctuations between the class-2
block-states, 
\begin{equation} 
\Psi_{i^*jkl^*} = \psi^L_{i^*} \otimes
\phi_j \otimes \phi_k \otimes \psi^R_{l^*} \equiv |i^*jkl^* \rangle
\end{equation} 
($m < i^*,l^* \leq m+p$) without increasing the
dimension of the Hilbert space. In analogy with the theory of
intermediate Hamiltonians~\cite{intham}, we intend to map the Hilbert
space ${\cal E}$ of the standard DMRG procedure ($m+p$ states per
block) onto the 3-classes DMRG Hilbert space ${\cal S}$ (with $m$
class-1 states and $p$ class-2 states). An effective Hamiltonian is
thereby defined acting on ${\cal S}$ as model space, which takes into
account the effects of the outer space ${\cal Q=E-S}$. 

Let us consider the secular equation of a state $\Psi_{abcd} \in {\cal S}$
\begin{equation}
\sum_{|ijkl\rangle \in {\cal S}} C_{ijkl} H_{ijkl,abcd} + 
\sum_{|\alpha\beta\gamma\delta\rangle \in {\cal Q}}
C_{\alpha\beta\gamma\delta}H_{\alpha\beta\gamma\delta,abcd} = 
C_{abcd} E 
\end{equation}
One would like to rewrite this equation in the following effective
Hamiltonian form
\begin{equation}
\sum_{|ijkl\rangle \in {\cal S}} C_{ijkl} H_{ijkl,abcd} + 
C_{abcd}\Delta_{abcd,abcd} = C_{abcd} E 
\end{equation}
since in this way the contribution of the states belonging to ${\cal Q}$ is
taken into account by a diagonal dressing $\Delta$ of the
Hamiltonian. $\Delta$ should therefore satisfy the equations
\begin{equation}
\Delta_{abcd,abcd} = \sum_{|\alpha\beta\gamma\delta\rangle \in {\cal Q}} 
{C_{\alpha\beta\gamma\delta} \over C_{abcd}}
H_{\alpha\beta\gamma\delta,abcd} 
\end{equation}

Notice however that the coefficients $C_{\alpha\beta\gamma\delta}$
cannot be obtained from the diagonalization of the effective
Hamiltonian $H_{eff} = H+\Delta$ since the latter acts only on the
model space ${\cal S}$. The main difficulty is therefore to
find a sound evaluation of the coefficients
$C_{\alpha\beta\gamma\delta}$. This problem can be solved using a
coupled-cluster approximation.

Let $S^L$ and $S^R$ be the cluster operators associated to the left
and right renormalized blocks and let $\Omega = \exp{(S^L+S^R)}$ be
the wave operator going from ${\cal S}_0$ to ${\cal E}$, where ${\cal
S}_0$ is generated by the class-1 blocks states only. ${\cal S}$ can
then be decomposed as ${\cal S = S}_0 + {\cal S}_R + {\cal S}_L$,
where ${\cal S}_R$ (${\cal S}_L$) is the Hilbert space generated
out of the product of class-2 states in the right (left)
renormalized block and class-1 states in the left (right)
renormalized block. The ground-state $\Psi \in {\cal E}$ can be expanded as
\begin{eqnarray}
\label{eq:psi1}
\Psi &=& 
   \sum_{|ijkl \rangle \in {\cal S}_0} C_{ijkl} \; |ijkl \rangle  + \\
&& \sum_{|i^*jkl \rangle \in {\cal S}_R} C_{i^*jkl} \; |i^*jkl \rangle +
    \sum_{|ijkl^* \rangle \in {\cal S}_L} C_{ijkl^*} \;|ijkl^* \rangle + \\
&& \sum_{|i^*jkl^* \rangle \in {\cal Q}} C_{i^*jkl^*} \;|i^*jkl^* \rangle \; , 
\end{eqnarray}
while the coupled-cluster approximation gives   
\begin{eqnarray}
\label{eq:psi2}
\Psi &=&    \sum_{|ijkl \rangle \in {\cal S}_0} C_{ijkl} \;|ijkl \rangle  + \\
&& \sum_{i^* \atop m<i^*\leq m\!+\!p} \sum_{|ijkl\rangle \in {\cal S}_0} 
       S^L_{i^*,i} C_{ijkl}\; |i^*jkl \rangle +
    \sum_{l^* \atop m<l^*\leq m\!+\!p} \sum_{|ijkl \rangle \in {\cal S}_0} 
       S^R_{l^*,l} C_{ijkl}\; |ijkl^* \rangle + \\
&& \sum_{i^*l^* \atop m<i^*\leq m\!+\!p,\;m<l^*\leq m\!+\!p}
       \sum_{|ijkl \rangle \in {\cal S}_0} 
       S^L_{i^*,i}S^R_{l^*,l} C_{ijkl}\; 
       |i^*jkl^* \rangle \; .
\end{eqnarray}
Comparing equations~\ref{eq:psi1} and~\ref{eq:psi2} one obtains the 
definition of the cluster operators 
\begin{eqnarray*}
C_{i^*jkl} &=&  \sum_{ijkl \atop i\leq m,\;l\leq m} 
       S^L_{i^*,i} C_{ijkl} \; , \\
C_{ijkl^*} &=&  \sum_{ijkl \atop i\leq m,\;l\leq m} 
       S^R_{l^*,l} C_{ijkl} 
\end{eqnarray*}
and the evaluation of the unknown coefficients
\begin{equation}
C_{i^*jkl^*} =  \sum_{il \atop i\leq m,\;l\leq m} 
       S^L_{i^*,i}S^R_{l^*,l} C_{ijkl}  \; .
\end{equation}
These systems of equations are unfortunately over-defined in most
practical cases ($p>m$). An optimal choice for $S^L$ is obtained by
minimizing  the square deviation 
\begin{eqnarray*}
{\cal L}^L &=& \sum_{i^*jkl \atop m<i\leq m\!+\!p,\;l\leq m} 
(C_{i^*jkl} - \sum_{i \atop i<m} S^L_{i^*,i} C_{ijkl})^2 \; .
\end{eqnarray*}
Thus, 
\begin{eqnarray*}
{\partial {\cal L}^L \over \partial S^L_{i^*,i}} &=& 
-2\left(C_{ijkl} C_{i^*jkl} - 
\sum_{i' \atop i'<m} S^L_{i^*,i'} C_{ijkl} C_{i'jkl} \right) \\
&=& -2\left( \rho^L(i^*,i) - \sum_{i' \atop i'<m} S^L_{i^*,i'} \;
\rho^L(i',i) \right) \; = \; 0 \; ,
\end{eqnarray*}
where $\rho^L$ refers to the ground-state density matrix reduced to
the left renormalized block. Finally, one obtains 
\begin{equation}
S^L = \left(P^L_2 \rho^L P^L_1\right) \left(P^L_1 \rho^L P^L_1\right)^{-1}
\end{equation}
where $P^L_1$ ($P^L_2$) is the projection operator onto the class-1
(class-2) left block-states. Similar equations hold for the right block.

It should be noted that the clusters operators depend on the
ground-state of the dressed Hamiltonian, which depends itself on
the clusters operators. The whole process should then be iterated up
to self-consistency.
Let us finally recall that self-consistent approximations are usually 
very efficient but do not necessarily satisfy the variational 
principle.

%%%%%        %%%%%
%%%%% TABLES %%%%%
%%%%%        %%%%%

\newpage
\begin{table}
\caption{
Ground-state energy $E$ of the infinite spin-$1/2$ Heisenberg chain 
as obtained using the 3-classes DMRG method with different numbers 
of block states $m$ and $p$ in class-1 and class-2 subspaces 
(see Fig.~\protect\ref{fig:st}). 
$E_{ex} = -J \ln 2 + J/4$ refers to the Bethe-ansatz exact result 
and $E_0$ to a standard DMRG calculation with $m = 21$ block states 
($p=0$). The results labeled $m=m+p=\infty$
are obtained by linear extrapolation of the truncation error
$\epsilon(m,p) = \Sigma_2^2 + 2\Sigma_3 \to 0$ using the 
results for $m=5$ and $m=7$. $\Delta\varepsilon = (E-E_0)/(E_0 - E_{ex})$ and
$D_0/D = (m+p)^2 / (m^2 + 2mp)$ indicate, respectively, the loss of accuracy
and the reduction of the dimension of the Hilbert space with respect to
standard DMRG ($m=21$).
        }
\begin{tabular}{cccccc}
$m$ &$m+p$ &$E-E_{ex}$ &$\Sigma_2^2 + 2\Sigma_3$ &$\Delta\varepsilon$ & $D_0/D$ \\
\tableline  
5   &21  &$1.07 \times 10^{-4}$   &$3.72\times 10^{-5}$  &3.19   &2.4 \\
7   &21  &$4.47 \times 10^{-5}$   &$1.24\times 10^{-5}$      & 0.75  &1.8 \\
10  &21  &$3.34 \times 10^{-5}$   &$0.70\times 10^{-5}$      & 0.31  &1.4 \\
14  &21  &$2.74 \times 10^{-5}$   &$0.43\times 10^{-5}$      & 0.076 &1.1 \\
21  &21  &$2.55 \times 10^{-5}$                                       \\
$\infty$ &$\infty$  &$1.37 \times 10^{-5}$                                       \\
\end{tabular} 
\label{tab:Heis}
\end{table}

%%%%%         %%%%%
%%%%% FIGURES %%%%%
%%%%%         %%%%%

\newpage

\vfill\break

\vfill\break

\end{document}